\documentstyle[12pt]{article}

\begin{document}
\bibliographystyle{unsrt}

\topmargin 0pt
\oddsidemargin 5mm
\begin{titlepage}
\setcounter{page}{0}
\rightline{Preprint YERPHY-1429(16)-94}

\vspace{2cm}
\begin{center}
{\Large "Pseudoscalar and vector correlators in instanton  vacuum
model"}
\vspace{1cm}

{\large S.V.Esaibegyan  and S.N.Tamaryan.} \\
\vspace{1cm}
{\em Yerevan Physics Institute, Alikhanyan Brother street 2.
375036 Yerevan 36, Armenia.}\\
\end{center}

\vspace{5mm}
\centerline{{\bf{Abstract}}}

It is shown, that the correct account of quark interaction
generated  by instanton medium  gives  the  possibility of
description on  the same ground  of both pseudoscalar  channel and
low lying bound states in the vector channel. It must be noted,
that all calculations are made in the approximation, taking into
account only zero modes.

\end{titlepage}
\newpage
\renewcommand{\thefootnote}{\arabic{footnote}}
\setcounter{footnote}{0}

In this work we should like to show,  that largescale
instanton fluctuations may be responsible not only for
description of the mechanism of spontaneous  breaking of chiral
symmetry (SBCS),but also can be crucial for bound
state generation in vector channel.

 The problem of understanding of the mechanism of SBCS is in the focus of
attention during the last decade (see for  e.g. \cite{EV}).There were
many attempts to find this solution both from the qualitative
\cite{EV1}-\cite{MA} and the quantitative \cite{DV}-\cite{SS}
points of view.The  main idea \cite{DV,DV1} was,that mixing and
collectivization of zero instanton $I$ and antiinstanton $\bar{I}$
modes can bring to SBCS. The model of vacuum-$I \bar{I}$ undensed
"liquid" was found. But the approches developed  in
\cite{DV1}-\cite{SS}, did not grasp such basic fact,
that the quarks are in a continuous medium and as a consequence of this,
exchange momenta.

The  physical picture was as follows: the medium was  considered
as a discrete totality of  $I \bar{I}$-pairs and  only in the last
stage of calculations the transition to the
continuous limit  was carried out.
It was correct, when quark propagation in
instanton vacuum  was considered. The medium induced
"breaking" effect, from which effective quark mass $M(p)$ arises,
does  not depend on the order of  transitions to termodynamic limit
($N\rightarrow\infty V\rightarrow\infty, N/V=const,$ N-number of
"pseudoparticles", V-4 volume). The same approach is applicable
at to the Goldstones  modes (pions) description,  where the  main role
has  the quarks interaction with  vacuum  and  the momentum exchange
between  quarks is not so essential (see below).

Obviously,  when  other particles are  considered, for example,
the vector particle, the continuity of  medium plays  an
essential role, as just through this medium quarks have momentum
exchange and can form the bound state.

The  problem  of correct  account  of both effects -  the  quark
interactions with large scale instanton fluctuation of vacuum and
quark interactions between them, may be solved by averaging of
the total QCD Lagrangian in  statistical ansamble of
pseudoparticles \cite{SS1}. In  this  point, our  approach
essentially differs from  the previous one \cite{DV}-\cite{EV3} where
the separate correlators were averaged.
The effective action  in   instanton medium has the following
form  (in  Euklidean space), we work in the chiral limit,when $ m\rightarrow
0$, and in the  leading approximation by $N_c\rightarrow\infty$.
\begin{eqnarray}
\label{SS}
S & = & \int\left[\psi^+[\hat{k}-iM(k)]\psi+\chi^+\hat{k}\chi\right]\frac{d^4
k}{(2\pi)^4}+\nonumber\\
  & + & 2i\frac{V}{N}\int[M(k)M(q)M(p)M(l)]^{1/2}\delta^4(k+p-q-l)\frac
{d^4 k d^4 p d^4 q d^4 l}{(2\pi)^{12}}\nonumber\\
& \times & \left[(\psi^+_L(k)\psi_L(q))(\chi_L^+
(p)\chi_L(l))-(\psi^+_L(k)\chi_L(q))(\chi^+_L(p))\psi_L(l))
\right]+(L\rightarrow R),
\end{eqnarray}
This expression differs from the similar one in \cite{SS1}(for
$\chi^+_L, \chi_L$) by presence of the first  member  in the last
bracket,  that is connected with the transitions to new variables
$\chi_L^{'+},\chi_L'$ which have zero vacuum average
\begin{equation}
\chi^{a+}_{Li}\chi^b_{Lj}=\chi^{'a+}_{Li}\chi^{'b}_{Lj}+\frac{1}{4
N_c}<\chi^+_L\chi_L>\delta_{ab}\delta_{ij}
\end{equation}
Bozons fields $\chi^{'+}_L,\chi'_L$ with spin 1/2 in
(\ref{SS}),are in fact, new variables,  in  which collective
variables are transformed  after averaging
\begin{equation}
\label{AA}
M(p)=\frac{(N\epsilon)}{2VN_c})a^2(p),\,\quad a(p)=|p|\Phi(p)
\end{equation}
$\Phi(k)$ -is  connected with Fourier-transformed zero modes and
has the following asymptotic behaviour \cite{DV1}
\begin{equation}
\label{SB}
\Phi(p)=\cases{-2\rho\pi/|p|\quad\quad p\rho<<1;\cr
-12\pi/p^4\rho^2\quad p\rho>>1.\cr},
\end{equation}
$\varepsilon$-is determined from the gap equation \cite{DV1}
\begin{equation}
\label{SD}
1=\frac{4VN_c}{N}\int\frac{d^4p}{(2\pi)^4}\quad\frac{M^2(p)}{p^2+M^2(p)}.
\end{equation}

The action (\ref{SS})  has two essential picularities \cite{SS1}:
a) the perturbative theory is applicable as the vertex function
is small and besides, the model is superconvergent
(see(\ref{AA}),(\ref{SB})),
b) the potential leaves the ghost fields without masses,  in
approximation leading by $1/N_c$. The keeping potential $\sim 1/N_c$
(remember that $N\sim N_c$)is correct since it brings to results
$\sim N_c$ for  the correlators of colorless  currents. ($1/N_c$ factor
compensated by degrees $N_c$ from  the fermion ghost loops). Two
different members in (\ref{SS}) have various physical natures.The first
member,  in which quarks and  ghost fields  are  in separate
multipliers, is discribing interaction with longrange vacuum
fluctuations (see below) and  is similar to the "longitudinal
force" - Stokes resistance, which arises in liquid (see for
ex.\cite{NS}).The  second nonfactorized member, allows to
discribe the momentum tranzition between quarks, and. as a
consequence, is responsible for  the bound states. It's
nature reminds the "transversal friction forces"  which
arise in viscous liquid at motion  of two solid bodies .\cite{NS}

First we will consider  correlators of pseudoscalar currents
and show that the Goldstone mode- $\pi$ meson can be described and the answer
coincides with the  results of the diagram approach \cite{EV2}.
This calculation, in fact,  is our "reference point"
once more  proves the correctness of results
(\ref{SS}) for the effective action  (see also \cite{SS1}).

Note, that the pion interpretation as the Goldstoun particles,
connected with SBCS, means the following physical pictiure:
vacuum is packed by quark-antiquark pairs in $^3P_0$ state. If same
external agent flips the spin of  pair,
$^3P_0\rightarrow$ $^1S_0$,then there emerges a  pseudoscalar state,  the
pion. Masslessness of the  pion implies that such spin alignment
costs  no energy, and therfore, the  nonperturbartive vacuum
fluctuations remain,  in a sense,untouched \cite{NS}.The
longitudional forces  in pseudoparticle "liquids" just bring to
such a  picture.
The correlation function   is defined as:
\begin{equation}
\label{CF}
\Pi^\Gamma(p)=-\int (dk_1 dk_2)J_\Gamma (k_1)J_\Gamma (k_2)e^{-S}
\end{equation}
$$
\int (dk_1 dk_2)\equiv\int \frac{d^4k_1\,d^4k_2 }{(2\pi)^4}\delta^4(k_1-k_2+p),
\quad \Gamma=1,\gamma_5,\gamma_\mu,\gamma_\mu\gamma_5,\sigma_{\mu\nu}
$$
In approximation leading by $N_c \Pi^5(p)$ are difined by the
graphics in fig.1. The transversal  forces  are dumped as
$1/N_c$ and we are calculating in the leading approximation
of medium packing parameter($\rho / R\simeq\frac{1}{3}$)\cite{DV}
(R is the mean destance between  pseudoparticles $R^{-1}\simeq 200{\rm MeV}$)
Let us  define the  vertex by fig.2 (all $N_c,\frac{N}{V}$we
shall  take to the  correlator). At $\Gamma=\gamma_5$ we have
$(M_i\equiv M(k_i))$
\begin{equation}
\label{CG}
\Gamma^5_{\pm}(p)=\int Sp\gamma_5(\hat{k_1}+iM_1)\frac{(1\pm\gamma_5)}{2}
(\hat{k_2}+iM_2)
\frac{\sqrt{M_1M_2}}{(k_1^2+M_1^2)(k_2^2+M_2^2)}\frac{d^4 k}{(2\pi)^4}
\stackrel{p\rightarrow 0}{\simeq}\pm\frac{<\bar{\psi}\psi>}{2N_c}
\end{equation}
It is easy  to  check  that  in the vector ($\Gamma=\gamma_\mu$) and tensor
channels $(\Gamma=\sigma_{\mu\nu})$ the vertex functions are zero
in  accordance with the previous results \cite{EV2},\cite{DV1}.
Using definition of  $\Gamma_{\pm}$ we can write the connected  part
of $\Pi^5$  in the  following symbolical form:
\begin{eqnarray}
\label{CH}
\Pi(p) & = & 2N_c\Gamma_{\pm}(p)\left\{
        \left(
        \frac{2VN_c}{N}
        \right)^2G+
                \left(\frac{2VN_c}{N}\right)^4
                GFG+\ldots \right\}
\Gamma_{\mp}(p)=\nonumber\\
       & = &\frac{4VN_c^2}{N}\Gamma_{\pm}(p)\frac{1}{R_{-}(p)}
       \Gamma_{\mp}(p),\nonumber\\
 \frac{1}{R_{-}(p)} & = &
\frac{G\frac{2VN_c}{N}}{1-\left(\frac{2VN_c}{N})\right)^2FG}
\end{eqnarray}
Here G is ghost loop contribution
\begin{eqnarray}
\label{CJ}
\frac{2VN_c G}{N}=\frac{2VN_c}{N}\int\frac{d^4u}{(2\pi)^4}
\left[Sp\frac{\hat{u}+\hat{p}}{(u+p)^2}\left(\frac{1\pm\gamma_5}{2}\right)
\frac{\hat{u}}{u^2}\left(\frac{1\mp\gamma_5}{2}\right)\right]\times
\nonumber\\
\times M(u)M(u+p)\stackrel{p\rightarrow 0}{\simeq}
\frac{4VN_c}{N}\int \frac{d^4u}{(2\pi)^4}\frac{M^2(u)}{u^2}=1+0(\rho/R),
\end{eqnarray}
where we used  the relations(\ref{SD}). F is fermion loop contribution
\begin{eqnarray}
\label{CK}
\frac{2VN_c}{N} F & = & \frac{2VN_c}{N}\int (dk_1dk_2)M_1M_2 Sp
\frac{(\hat{k}_1+iM_1)}{k_1^2+M_1^2}\left(
\frac{1\pm\gamma_5}{2}\right)\frac{\hat{k}_2+iM_2}{k_2^2+M_2^2}
\left(\frac{1\mp\gamma_5}{2}\right)=\nonumber\\
& = & \frac{4VN_c}{N}\int(dk_1dk_2)M_1M_2\frac{k_1 k_2}{(k_1^2+M_1^2)
(k_2^2+M_2^2)}+O(\rho/R)
\end{eqnarray}
Using (\ref{SD})and (\ref{CJ})(\ref{CK}) for $R_(p)$ we have :
\begin{eqnarray}
\label{CL}
R_-(p)=\frac{2VN_c}{N}\int(dk_1dk_2)\frac{(M_1k_{2\mu}-M_2k_
{1\mu})^2}{(k_1^2+M_1^2)(k_2^2+M_2^2)}\stackrel{p\rightarrow
0}{\simeq}p^2\frac{2VN_c}{N}\int
\frac{d^4k}{(2\pi)^4}\frac{M^2}{(k^2+M^2)^2}\simeq \beta p^2.
\end{eqnarray}
Collecting (\ref{CG}) and (\ref{CL}) we obtain:
$$ \Pi^5(p)=N_c
\frac{VN_c}{N}\left[\frac{<\bar{\psi}\psi>}{N_c}\right]^2\frac{1}{\beta
p^2} $$
It is remarkable  that the connected part (\ref{CH}) has a  resonanse
form; the zeroes of $ R_{-}(p)  $ determine  the position of  the
poles in  $p^2$. Using the well-known asymptotic relation,
following from the current algebra
$$ \Pi^5(p)\stackrel{p\rightarrow 0}{\simeq}\frac{4<\bar{\psi}\psi>^2}
{f^2_\pip^2} $$
We finally obtain for $ f_\pi $
\begin{equation}
\label{CM}
f_\pi^2=4\beta\frac{N}{V}=8N_c \int\frac{d^4k}{(2\pi)^4}\frac{M^2}{(k^2+M^2)^2}
\simeq \frac{N_c}{\pi^2}M^2(0)\ln\frac{1}{M(0)\rho},
\end{equation}
(numerically $ f_\pi=142{\rm MeV}$) which coincides with the diagram results
\cite{DV1}.

Let us turn now to the vector channel. In fig.3 we show the  diagram
types which arise when the transvers forces are taken into
account ("Longitudinal forces" are dumping as $1/N_c$ in this
case). Our goal, however,  is the extraction of one  meson states.
So if we  require that the cutting of any diagram
contains only two  quarks, and  ghost ends must not form white
object,  then all the diagrams of  fig.3c may be
rejected (remember the general rule at $N_c\rightarrow\infty$:
diagrams are planar and the number
of quark loops is minimal \cite{NS}). Acting so we, in fact, neglect
the contributions from continiuum in correlator. So the connected part
of correlator is defined by diagrams of fig.3a,b, which we can
sum using the Fredgolm equation by standard way
\begin{eqnarray}
\label{SF}
\Pi_{\mu\nu}(_p) & = &-\int(dk_1dk_2)j_\mu(k_1)j_\nu(k_2) e^{-S}
 =
2N_c\lambda\int\Gamma_\mu(u,p)\Gamma_\nu^+(u,p)\frac{d^4u}{(2\pi)^4}\nonumber\\
&  & +2N_c\lambda^2\int\Gamma_\mu(u,p)R(u,v)
\Gamma_\nu^+(v,p)\frac{d^4ud^4v}{(2\pi)^8}
\end{eqnarray}
where $\lambda=\frac{4V^2N_c}{N^2}$; vertex $\Gamma_\mu$ is
define on fig.3 by shaded area:
\begin{eqnarray}
\label{SG}
\Gamma_{\mu\pm}(u,p)=\int
Sp\left(\frac{(\hat{k}+iM)}{k^2+M^2}\gamma_\mu
\frac{(\hat{k}+\hat{p}+iM)}{(k+p)^2+M^2}\frac{(1\mp\gamma_5)}{2}
\frac{(\hat{u}+\hat{k})}{(u+k)^2}\frac{(1\pm\gamma_5)}{2}
\right)\times\nonumber\\
\times\sqrt{M(k)M(k+p)}\quad M(u+k)\frac{d^4k}{(2\pi)^4},\quad \quad
\Gamma_{\mu+}=\Gamma_{\mu-}=\Gamma_\mu
\end{eqnarray}
$R(u,v)$ satisfies the Fredgolm equation
\begin{equation}
\label{SH}
R(u,v)=S(u,v)+\lambda\int S(u,k)R(k,v)\frac{d^4k}{(2\pi)^4},
\end{equation}
where kernel  $S(u,v)$is represented in fig.4
\begin{eqnarray}
\label{SJ}
S_1\simeq\int & & Sp \left(\frac{\hat{k}}{(k^2+M^2)}\frac{(1\pm\gamma_5)}{2}
\frac{(\hat{u}+\hat{k})}{(u+k)^2}\frac{(1\mp\gamma_5)}{2}\frac{(\hat{k}
+\hat{p})}{(k+p)^2+M^2}\frac{(1\pm\gamma_5)}{2}
\frac{(\hat{v}+\hat{k})}{(v+k)^2}\frac{(1\mp\gamma_5)}{2}
\right)\nonumber\\
& & \times M(k)M(k+p) M(u+k)M(v+k)\frac{d^4k}{(2\pi)^4}\nonumber\\
S_2\simeq N_c &  & \left(\int
M(k)M(u+k)Sp\frac{\hat{k}}{k^2+M^2}\frac{(1\pm\gamma_5)}
{2}\frac{(\hat{u}+\hat{k})}{(u+k)^2}\frac{(1\mp\gamma_5)}{2}
\frac{d^4k}{(2\pi)^4}\right)^2
\end{eqnarray}
Using i)that in leading $\rho/R$approximations  $\rho M\ll
1,\rho u\ll 1,\rho v \ll 1$; ii)the scale at which $M(k)$ is
changed is $1/\rho$ see (\ref{AA}),we exchange in $S_1
M(k_i)\rightarrow M(0)$ and cut the integral on upper limit by
$  k  \leq 1/\rho $ \cite{DV1}. When  external particle momemtum
$p_\mu\rightarrow 0 $ we obtain
\begin{eqnarray}
\label{KA}
&  & S(u-v)\stackrel{p\rightarrow 0}{\simeq}\frac{M^4(0)}{8\pi^2}\ln
[\rho^2(u-v)^2]+\frac{1}{\lambda}(2\pi)^4 \delta^4(u-v)\nonumber\\
&  & \Gamma_\mu(u,p)\stackrel{p\rightarrow0}{\simeq}\frac{M^2(0)}{16\pi^2}
\left[(2u_\mu-p_\mu)\ln(\rho^2u^2+\rho^2M^2)+\frac{2}{3}p_\mu\frac{pu}{u^2}
\ln(\frac{M^2}{u^2+M^2})\right]
\end{eqnarray}
As kernel $S$ is a function of the difference $(u-v)$, equations
(\ref{SH})can be solved by Fourier-transformation
\begin{equation}
\label{KB}
R(x)=P\frac{S(x)}{1-\lambda S(x)},
\end{equation}
where P denotes the  principal value.
\begin{equation}
\label{KC}
\Pi_{\mu\nu}(p)=2N_c\lambda\int
\Gamma_\mu(p,x)\Gamma_\nu^+(p,x)d^4x+2N_c\lambda^2 P\int
\Gamma_\mu(p,x)\frac{S(x)}{1-\lambda S(x)}\Gamma_\nu^+(p,x)d^4x
\end{equation}
Using (\ref{KA}), we find at $p\rightarrow0$ following
expression for correlator in vector channel:
\begin{eqnarray}
\label{KD}
\Pi_{\mu\nu}(p)  & = & 2N_c\lambda P\int
\frac{\Gamma_\mu(p,x)\Gamma_\nu^+(p,x)}{1-\lambda S(x)}d^4x=-2N_c\lambda
P\int\frac{\Gamma_\mu(p,x)\Gamma_\nu^+(p,x)}{S_1(x)}d^4x\nonumber\\
& =
&-\frac{N_c}{16\rho^2\pi^2}\left[-I\delta_{\mu\nu}+\frac{8}{3}\rho^2(\ln\rho^2M^2)^2
p_\mu p_\nu\right]\nonumber\\
I& = &P\int\limits_{0}^{\infty}\frac{zJ^2_2(z)dz}{2J_0(z)+zJ_1(z)-2}\simeq -5.5
\end{eqnarray}
Here $J_i(z)$ are Bessel functions of first kind. We calculate I
numerically.
Correlator in vector channel,generally speaking,are
transversal. However,we have neglected the contributions from
continuum, and now we have only  the contribution  from lowlying
bound states. One may also show \cite{IZ},that in the $p\rightarrow
0$ limit (corresponding to
$x\rightarrow \infty$) the main  contributions in
Chellen-Leman represantation are defined from  one particle
states, and the correlator in our case (in Euclidean space) has
the  following form \cite{KLZ}:
\begin{equation}
\label{KN}
\Pi_{\mu\nu}(p)=-\Sigma \left(\delta_{\mu\nu}+\frac{p_\mu
p_\nu}{m_i^2}\right)f_i^2
\end{equation}
We have defined the current matrix  element as follows:
$<0|\bar{d}\gamma_\mu u|\rho>=e_\mu^\lambda f_\rho  m_\rho ;
f_{\rho  exp}\approx 200{\rm MeV}, e_\mu -\rho$ -meson polarization vector.If
later the smallness of $f_i$ for excited $(\rho'\ldots)$ states
\cite{GI}is taken into account, then we can  finally write
\begin{equation}
\label{KM}
\Pi_{\mu\nu}(p)=-\left(\delta_{\mu\nu}+\frac{p_\mu
p_\nu}{m_\rho^2}\right)f_\rho^2
\end{equation}
Comparing this with eq.(\ref{KD}), we obtain
\begin{equation}
\label{KL}
f_\rho^2=\frac{5.5N_c}{16\rho^2\pi^2}\quad \quad
m_\rho^2=\frac{3\times5.5}{8\rho^2(\ln\rho^2M^2)^2}
\end{equation}

We see that constant $f_\rho$ and $m_\rho$ are defined by the
values of pseudoparticles average size $(1/\rho=600{\rm
MeV})f_\rho=193{\rm MeV},m_\rho=797{\rm MeV}$ and  are in a good
agreement with the experemental data. But it should be noted, however,
that the  approximation which we used, gives $\sim30\%$  errors in
estimation, we should not like therefore to stress an
exceptional numerical agreement with experemental data. But
let's note again, that the effective action (\ref{SS}) makes it
possible to describe physical particles, and the vector
channel is described on  the same  ground as the pseudoscalar
one. In fact, our approach differs  from those
considered earlier  \cite{DV}-\cite{EV3}, by taking into
account the momemtum transitions owing to  viscous of instanton
"liquids". Vector channel in our approach  has  non  zero value
because of permanent scattering of quarks on
continious medium.

We are grateful  to I.G.Aznauryan, G.V.Grigoryan, A.G.Sedrakyan for helpful
discussons.The research described in this publication was made
possible in part by Grant \# RYEOOO from the International Science
Foundation and by INTAS Grant \# 93-183

\newpage
\begin{center}
Figure caption
\end{center}

Fig.1  Connected part of correlation function in pseudoscalar channel.
       Dashed lines correspond to ghost fields and solid lines
       quarks fields.
       Black points are $\frac{1\pm\gamma_5}{2}$

Fig.2  Vertex function in pseudoscalar  channel.

Fig.3(a,b,c) Types of graphs in vector channel. Shaded area is $\Gamma$.

Fig.4(a,b) Kernels $S_1-(a)$ and $S_2-(b)$.

\end{document}